\documentclass[aps,prb,twocolumn,showpacs,preprintnumbers,amsmath,amssymb,floatfix]{revtex4-1}

\usepackage{graphicx}
\usepackage{dcolumn}
\usepackage{bm}
\usepackage{CJK}

\begin{document}
\begin{CJK*}{GBK}{}

\title{Quantum thermal transport through anharmonic systems: A self-consistent approach}

\author{Dahai He}\email[]{dhe@xmu.edu.cn}
\affiliation{Department of Physics and Institute of Theoretical Physics and Astrophysics, Xiamen University, Xiamen 361005, China}
\author{Juzar Thingna}\email[]{juzar@smart.mit.edu}
\altaffiliation{Present address: Singapore-MIT Alliance for Research and Technology (SMART) Centre, Singapore 138602}
\affiliation{Physics Department, 2 Science Drive 3, National University of Singapore, Singapore 117551, Republic of Singapore}
\author{Jian-Sheng Wang}
\affiliation{Physics Department, 2 Science Drive 3, National University of Singapore, Singapore 117551, Republic of Singapore}
\author{Baowen Li}
\affiliation{Department of Mechanical Engineering, University of Colorado, Boulder, CO 80309}

\date{\today}

\begin{abstract}
We propose a feasible and effective approach to study quantum thermal transport through anharmonic systems. The main idea is to obtain an {\it effective} harmonic Hamiltonian for the anharmonic system by applying the self-consistent phonon theory. Using the effective harmonic Hamiltonian we study thermal transport within the framework of nonequilibrium Green's function method using the celebrated Caroli formula. We corroborate our quantum self-consistent approach using the quantum master equation that can deal with anharmonicity exactly, but is limited to the weak system-bath coupling regime. Finally, in order demonstrate its strength we apply the quantum self-consistent approach to study thermal rectification in a weakly coupled two segment anharmonic system.
\end{abstract}

\pacs{05.70.Ln, 44.10.+i, 05.60.-k}
\maketitle
\end{CJK*}
\section{Introduction}\label{sec:1}
Developing a first-principle based approach for quantum thermal transport across low-dimensional systems not only provides insight to potential nanodevice applications, but is also crucial to better understand nonequilibrium statistical physics. Till date, quantum thermal transport across harmonic crystals has been extensively studied using the generalized Langevin approach~\cite{Dhar06, Dhar_rev}, nonequilibrium Green's function method~\cite{Wang_revEPJB, Wang_revFront}, or the density matrix approach \cite{Dhar12}. The main advantage of these methods lies in their exactness of treating harmonic systems, giving rise to ballistic transport.

On the other hand, recent progress in classical thermal transport has demonstrated interesting practical applications, such as thermal diode, thermal transistor, and thermal logic gates~\cite{Li_rev}. These studies suggest that the ability to manipulate thermal transport may lead to important technological breakthrough ranging from novel devices to improvement of thermal management in microelectronics, and even information processing by phonons. Unfortunately, the exactly solvable harmonic crystals do not exhibit these novel properties and it turns out that anharmonicity is one of the key ingredients for their occurrence.

Naturally, anharmonic systems are of great interest in order to deduce the basic microscopic origin of these novel properties. Hence in the classical regime the effective phonon theory \cite{Nianbei06} and the self-consistent phonon theory \cite{Hu_SCPT06} were developed to study thermal transport for highly anharmonic systems. The former was based on the equipartition theorem, whereas the latter took advantage of the Feynman-Jensen inequality. In the quantum regime, the role of anharmonicity is even more enticing but relatively unexplored. A plethora of techniques for weakly anharmonic systems have been developed in this regime \cite{Segal03, Wang_revEPJB}, but a robust theory for strong anharmonicity still eludes the community. One of the popular techniques to treat the strongly anharmonic quantum regime is based on the quantum master equation that treats the system-bath interaction perturbatively \cite{Segal05, Thingna12, Thingna14}. Despite its ability to treat anharmonicity exactly, one major disadvantage is its inability to treat large phononic systems. This is mainly because this method operates in the eigenbasis of the system that increases rapidly with temperature or number of particles. Hence, an approach that can deal with relatively large systems is critical to understand the role of strong anharmonicity in thermal transport.

In this paper we will propose a feasible and effective approach to study thermal transport through anharmonic systems. The key idea here is to renormalize the anharmonic Hamiltonian to an effective harmonic Hamiltonian using the quantum self-consistent phonon theory \cite{He08a}. We then apply the standard nonequilibrium Green's function machinery to study the effective harmonic model. The paper is organized as follows: In Sec.~\ref{sec:2}, we introduce the anharmonic model and propose a modified Caroli formula for thermal transport based on the quantum self-consistent phonon theory. We then corroborate our quantum self-consistent approach with the help of quantum master equation for mono- and di-atomic molecular junctions in Sec.~\ref{sec:3}. Finally, we demonstrate an intriguing application of our method by investigating thermal rectification in Sec.~\ref{sec:4}. Finally, we summarize our main conclusions in Sec.~\ref{sec:5}.

\section{Quantum self-consistent phonon theory and modified Caroli formula}\label{sec:2}
\begin{figure}
\includegraphics[width=\columnwidth]{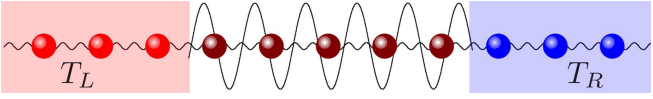}
\caption{\label{fig_model}(Color online) Schematic illustration of the anharmonic model given by Eq.~\eqref{s2-eq-Hamiltonian}. The left and right harmonic baths are at temperatures $T_L$ and $T_R$ respectively. The central system consists of harmonic plus anharmonic interactions depicted by the periodic potential}
\end{figure}
We consider the minimal model for thermal transport (as illustrated in Fig.~\ref{fig_model}) that consists of a general one-dimensional system linearly coupled to two semi-infinite chain of harmonic oscillators, herein referred to as heat baths. The corresponding Hamiltonian $H$ of the total system reads \cite{Weiss08},
\begin{eqnarray}\label{s2-eq-Hamiltonian}
H = H_{S} &+& \sum_{l}\frac{P_{l}^2}{2M_{l}}
+\frac{M_{l}\omega_l^2}{2}\left(Q_l-\frac{c_l S_{L}}{M_{l}\omega_l^2}\right)^{2} \nonumber\\
&+&\sum_{r}\frac{P_{r}^2}{2M_{r}}+\frac{M_{r}\omega_r^2}{2}\left(Q_r-\frac{c_r S_{R}}{M_{r}\omega_r^2}\right)^{2},
\end{eqnarray}
where $H_{S}$ describes the system of interest, \{$Q_{x}$, $P_{x}$, $M_{x}$, $\omega_{x}$\} are the positions, conjugate momenta, masses, and frequency modes of the left ($x=l$) and right ($x=r$) bath. The parameter $c_{x}$ is the system-bath coupling constant of the $x$-th mode corresponding to the left ($x=l$) and right ($x=r$) bath. The system operator $S_{\alpha}$ couples the system to the $\alpha$-th bath and in general it can be any system operator or its function. The above Hamiltonian is commonly referred to as the Zwanzig-Caldeira-Legett model \cite{Zwanzig73, Caldeira83} that can be split into various regions as,
\begin{equation}
H=H_{S}+H_{L}+H_{R}+\sum_{\alpha = L,R} H_{S\alpha}+H_{\alpha}^{RN}.
\end{equation}
The bath Hamiltonian
\begin{equation}\label{s2-eq-Hamiltonianbath}
H_{\alpha}=\sum_{x}\frac{P_{x}^{2}}{2M_{x}}+\frac{\omega_{x}^{2}}{2}Q_{x}^{2}.
\end{equation}
The system-bath interaction Hamiltonian
\begin{equation}\label{s2-eq-Hamiltoniansysbath}
H_{S\alpha}=S_{\alpha}\otimes B_{\alpha},
\end{equation}
where $B_{\alpha}=-\sum_{x}c_{x}Q_{x}$ is the collective bath operator that couples with the system and
\begin{equation}\label{s2-eq-Hamiltonianrn}
H_{\alpha}^{RN}=\frac{S_{\alpha}^2}{2}\sum_{x}\frac{c_{x}^{2}}{M_{x}\omega_{x}^{2}}
\end{equation}
is known as the re-normalization (counter) term. In this work since we will couple the system to the bath via the position-position coupling the re-normalization term is essential to maintain translational invariance of the total system. In the above equations $\alpha = L;~x=l$ corresponds to the left bath and $\alpha = R;~x=r$ corresponds to the right bath. In order to simplify the description, we have considered a one dimensional model, but our theory described below can be easily generalized to higher dimensions. The main difficulty to calculate heat current using the above Hamiltonian lies in the anharmonic interactions present in the system. To tackle such anharmonic interactions we apply the quantum self consistent phonon theory (QSCPT) to obtain an effective harmonic system \cite{Feynman86, He08a} and then obtain the heat current using the machinery of nonequilibrium Green's function (NEGF) \cite{Wang_revEPJB} through the effective model.

Without loss of generality, we consider the system as a one-dimensional oscillator chain of the form
\begin{equation}
H_{S}=\sum_{s}\frac{m}{2}\dot{x}_{s}^{2}+W(x_{s}-x_{s-1})+V(x_s),
\end{equation}
where $W(\delta x)$ and $V(x)$ are the nearest-neighbor interaction and onsite potential respectively. The partition function in the canonical ensemble can be expressed as a path integral over all possible trajectories, i.e.,
\begin{equation}
Z=\int\mathrm{D}\mathbf{x}e^{-\frac{S[\mathbf{x}]}{\hbar}},
\end{equation}
where the measure of functional integral $\mathrm{D}\mathbf{x} \equiv \Pi d\mathbf{x}$ and
\begin{equation}
\label{s2-eq-action}
S[\mathbf{x}]=\int_0^{\hbar\beta}d\tau\left(\frac{m}{2}\dot{\mathbf{x}}^2+W(\delta \mathbf{x})+V[\mathbf{x}]\right).
\end{equation}
In the action $S[\mathbf{x}]$ above, $\mathbf{x}$ and $\dot{\mathbf{x}}$ are implicit functions of the time variable $\tau$. The key idea of quantum self-consistent phonon theory (QSCPT) is to replace the original Euclidean action, Eq.~(\ref{s2-eq-action}), by an approximate one. In order to do this we make a reasonable choice of the trial Hamiltonian
\begin{equation}
\label{s2-eq-Hamiltonianeff}
H_{S}^{eff}=\sum_{s}\frac{m}{2}\dot{x}_{s}^{2}+\frac{f_{c}}{2} (x_{s+1}- x_{s})^{2}+\frac{f}{2}x^{2}_{s},
\end{equation}
where the parameters $f_c$ and $f$ are to be deduced by minimizing the right hand side of the Feynman-Jensen inequality \cite{Feynman98}:
\begin{equation}\label{s2-eq-feynman}
F\leq F_{0}+\langle H_{S}-H_{S}^{eff}\rangle_{\mathrm{\scriptscriptstyle{canonical}}},
\end{equation}
where $F_0=-k_{B}T\ln{Z_0}$. The trial partition function
\begin{equation}
Z_{0}=\int \mathrm{D}\mathbf{x}e^{-\frac{S[\mathbf{x}]}{\hbar}},
\end{equation}
where $S_{0}[\mathbf{x}] = \int_0^{\hbar\beta}d\tau\left(m\dot{\mathbf{x}}^2+f_{c}\delta \mathbf{x}^{2}+f\mathbf{x}^{2}\right)/2$. The canonical average in Eq.~(\ref{s2-eq-feynman}) is computed based on the trial Hamiltonian Eq.~(\ref{s2-eq-Hamiltonianeff}) that can be easily calculated since the integrand takes a quadratic form. Finally, the parameters $f_c$ and $f$ can be obtained by solving the following self-consistent equations:
\begin{eqnarray}
\label{eq:wp}
\omega_{p}^{2}&=&\frac{2}{m}\left[\frac{\partial V(\rho)}{\partial \rho^{2}}+4\frac{\partial W(\delta\rho)}{\partial (\delta\rho^{2})}\sin^{2}\left(\frac{p\pi}{N}\right)\right],\\
\label{eq:rho}
\rho^{2}&=&\langle x_{k}^{2}\rangle \nonumber \\
&=&\frac{\hbar}{2Nm}\sum_{p}\frac{1}{\omega_{p}}\coth\left(\frac{ \beta\hbar\omega_{p}}{2}\right),\\
\label{eq:drho}
\delta\rho^{2}&\equiv&\langle(x_{k}-x_{k-1})^{2}\rangle \nonumber \\
&=& \frac{\hbar}{2Nm}\sum_{p}
\frac{4\sin^{2}(p\pi/N)}{\omega_{p}}\coth\left(\frac{\beta\hbar\omega_{p}}{2}\right),
\end{eqnarray}
where $\beta=(k_{B}T)^{-1}$, $T=(T_{L}+T_{R})/2$ and the variables $\omega_p,~\rho,$ and $\delta \rho$ implicitly depend on $f_c$ and $f$. Note that the canonical average is performed at the average temperature $T$, which requires that the heat baths have minimal influence on the system. This assumption could be accomplished in a variety of scenarios, e.g., when the system-bath coupling is weak and the temperature difference is small or when the system is comprised of various segments each strongly interacting with its own bath and weakly interacting with each other. It is worth noting here that although the effective phonon theory \cite{Nianbei06} is similar to the quantum self-consistent phonon theory (QSCPT) described above, it does not capture the essential quantum physics since it relies on the validity of the equipartition theorem. This serves as our main motivation to use the QSCPT and study \textit{quantum} thermal transport.

Therefore, given the effective Hamiltonian the model described by Eq.~(\ref{s2-eq-Hamiltonian}) is approximated as,
\begin{equation}
H \approx H^{eff}_{S}+H_{L}+H_{R}+\sum_{\alpha =L,R} H_{S\alpha}+H_{\alpha}^{RN}.
\end{equation}
Using the standard techniques to treat harmonic systems \cite{Dhar_rev, Wang_revEPJB} we obtain the steady state heat current given by the Landauer-like formula as,
\begin{equation}\label{s2-eq-current}
I_{L}=-I_{R}=\frac{1}{2\pi}\int_{0}^{\infty} d\omega\hbar\omega \widetilde{T}(\omega)(f_{L}-f_{R}).
\end{equation}
The above formula is valid for any temperature difference between the left and right baths. The transmission function $\widetilde{T}(\omega)$ is given by a modified Caroli formula,
\begin{equation}
\widetilde{T}(\omega)=\mathrm{Tr}(G^{r}\Gamma_{L}G^{a}\Gamma_{R}),
\label{eq:caroli}
\end{equation}
where
\begin{eqnarray}
\label{s2-eq-Greensfunc}
G^r&=&\left[ m\omega^2I-\widetilde{K}-\Sigma_{L}^{r}-\Sigma_{R}^{r}\right]^{-1},\\
\label{s2-eq-Gamma}
\Gamma_{\alpha}&=&-2\mathrm{Im}(\Sigma_{\alpha}^{r}),\\
f_{\alpha}&=&\frac{1}{e^{\hbar\omega/(k_{B}T_{\alpha})}-1}.
\end{eqnarray}
Here $I$ is the identity matrix and $\Sigma_{\alpha}^{r}$ ($\omega$ dependence suppressed) is known as the retarded self-energy of the $\alpha$-th bath that completely depends on the properties of the bath. The effective force matrix $\widetilde{K}$ above is tridiagonal and $G^a=(G^r)^{\dag}$. It is important to note that the transmission function $\widetilde{T}(\omega)$ here is temperature dependent for anharmonic systems, since the trial parameters $f_c$ and $f$ are temperature dependent. Hence, owing to this inherent temperature dependence due to the self-consistent equations (\ref{eq:wp}, \ref{eq:rho}, and \ref{eq:drho}) we herein term Eq.~(\ref{eq:caroli}) as the modified Caroli formula. Such a temperature dependent transmission function has been observed previously in the context of mean-field approximations \cite{Thingna12, Zhang13, Li13}. It is worth emphasizing that QSCPT \cite{He08a} has previously been used to evaluate the thermal conductivity via the kinetic theory as $\kappa = C v l$ with $C$ being the heat capacity, $v$ the phonon velocity, and $l$ being the mean free path \cite{Shrivastava}. It is in this work that we for the first time integrate the QSCPT with NEGF giving the ``quantum self-consistent approach'' a firm theoretical basis to be applied in the nonequilibrium \textit{quantum} regime.

\section{Corroborating the quantum self-consistent approach}\label{sec:3}
The theory outlined above is a general approach applicable to any anharmonic system that can be approximated as a harmonic one. In this section, we will take specific examples of the system Hamiltonian $H_{S}$, i.e., a monoatomic molecule confined in a quartic potential and a diatomic molecule with a quartic interaction and/or onsite potential. The concerned system is linearly connected to two heat baths via the position operator, i.e.,
$S_{L}=S_{R}=x$ for monatomic molecule and $S_{L}=x_1,S_{R}=x_2$ for the diatomic case. In order to specify all properties of the baths we define a spectral density
\begin{equation}
J_{\alpha}(\omega)  = \frac{\pi}{2} \sum_{x} \frac{c_{x}^2}{M_{x}\omega_{x}} \delta\left(\omega - \omega_{x}\right),
\end{equation}
that incorporates the effect of system-bath coupling since it is proportional to $c_{x}^{2}$. Above $\alpha = L;~ x=l$ corresponds to the left bath and $\alpha = R;~ x=r$ to the right. Given the above definition we can now recast the re-normalization part of the Hamiltonian Eq.~(\ref{s2-eq-Hamiltonianrn}) as,
\begin{equation}
H_{\alpha}^{RN} = \frac{S_{\alpha}^{2}}{\pi} \int_{0}^{\infty}d\omega\frac{J_{\alpha}(\omega)}{\omega} = \frac{S_{\alpha}^2}{2}\gamma_{\alpha}(0),
\end{equation}
where $\gamma_{\alpha}(0)$ is commonly referred to as the damping kernel at time $0$ of the $\alpha$-th bath. The self-energy $\Sigma_{\alpha}^{r}$ used in Eq.~(\ref{s2-eq-Greensfunc}) can now be expressed in terms of the spectral density \cite{Saito07} as,
\begin{equation}
\Sigma_{\alpha}^{r} = \frac{1}{\pi}\mathrm{P}\int_{-\infty}^{+\infty}\frac{J_{\alpha}(\omega^{\prime})}{\omega-\omega^{\prime}}d\omega^{\prime}+\gamma_{\alpha}(0) - i J_{\alpha}(\omega),
\end{equation}
where $\mathrm{P}$ denotes the principal value integral and importantly we have added the term coming from the re-normalization part of the Hamiltonian, i.e., $\gamma_{\alpha}(0)$, to the self-energy. It is important to note here that if the system-bath coupling has a nonlinear form then the re-normalization term should be incorporated in the effective Hamiltonian Eq.~(\ref{s2-eq-Hamiltonianeff}).

Throughout this work we will choose both the left and right baths to have the same properties, i.e., $J_{L}(\omega) = J_{R}(\omega) = J(\omega)$ and use a specific form of this spectral density, namely,
\begin{equation}
J(\omega)=\frac{\gamma m \omega}{1+(\omega/\omega_{c})^{2}},
\end{equation}
where the parameter $\gamma$ is the Stokesian damping coefficient and characterizes the system-bath coupling strength. The above spectral density corresponds to the Lorentz-Drude model of the heat bath, i.e., an Ohmic bath with a Lorentz-Drude cutoff $\omega_c$. Note that the theory given in Sec.~\ref{sec:2} is not restricted to any particular form of the spectral density, but the Lorentz-Drude form helps us to evaluate $H_{\alpha}^{RN}$ and $\Sigma^{r}=\Sigma_{L}^{r}+\Sigma_{R}^{r}$ analytically as,
\begin{eqnarray}
H_{\alpha}^{RN}&=&\frac{S_{\alpha}^{2}}{2}\gamma m \omega_{c} ,\\
\Sigma^{r}&=&2J(\omega)\left[\frac{\omega}{\omega_{c}}-i\right],
\end{eqnarray}
which immediately leads to $\Gamma_{\alpha}=-2\mathrm{Im}(\Sigma_{\alpha}^{r})=2J(\omega)$ via Eq.~(\ref{s2-eq-Gamma}).

\subsection{Monoatomic molecule}
The Hamiltonian for the monoatomic molecule is given by
\begin{equation}\label{s3-eq-hamiltonian}
H_{S}=\frac{p^{2}}{2m}+\frac{k}{2}x^{2}+\frac{\lambda}{4}x^{4},
\end{equation}
where $k$ and $\lambda$ are the spring constant and anharmonic strength for the quartic potential. According to QSCPT, we can obtain the effective Hamiltonian for this model as,
\begin{equation}
H_{S}^{eff}=\frac{p^{2}}{2m}+\frac{f}{2}x^{2},
\end{equation}
where the effective force constant $f$ is obtained by solving the self-consistent nonlinear equation;
\begin{equation}\label{s3-eq-f}
f=k+\frac{3\hbar\lambda}{2m\Omega}\coth\left(\frac{\beta\hbar\Omega}{2}\right),
\end{equation}
where $\Omega=\sqrt{f/m}$. Using the Landauer-like formula Eq.~(\ref{s2-eq-current}) one can obtain the heat current for the monoatomic molecule as
\begin{equation}\label{s3-eq-current-monatomic}
I_{L}=\frac{1}{2\pi}\int_{0}^{\infty}d\omega\hbar\omega G^{r}\Gamma_{L} G^{a}\Gamma_{R}(f_{L}-f_{R}),
\end{equation}
where $G^{r}$, $G^{a}$, $\Gamma_{L}$ and $\Gamma_{R}$ are all numbers for a single degree of freedom with $G^{r}(\omega)=[\omega^{2}-f-\Sigma^{r}(\omega)]^{-1}$.
\begin{figure}
\includegraphics[width=\columnwidth]{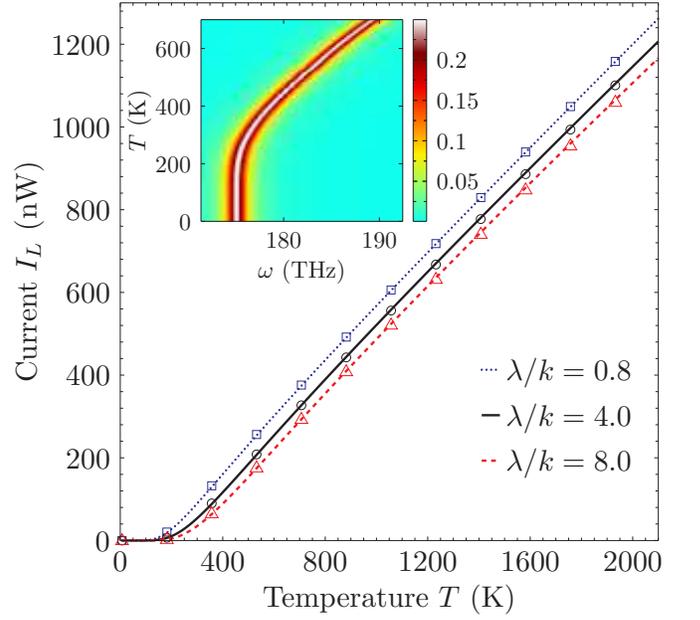}
\caption{\label{fig1}(Color online) Current $I_{L}$ as a function of temperature $T$ for various strengths of quartic potential in a monoatomic molecule. Lines correspond to the quantum self-consistent approach and the empty symbols correspond to quantum master equation. Inset shows the temperature dependence of the transmission function for $\lambda/k=8$. All common parameters are: $m=$ 1u, $k=$ 30.160meV/($\textup{\AA}^2$u), $\gamma=$ 0.92THz, $\omega_{c}=$ 0.92PHz, and $\Delta=0.05$. The left bath temperature $T_{L} = T\left(1+\Delta\right)$, whereas the right bath temperature $T_{R} = T\left(1-\Delta\right)$. $\lambda/k$ has dimensions of [$\textup{\AA}^2$u]$^{-1}$.}
\end{figure}
Figure~\ref{fig1} shows the heat current $I_{L}$ calculated via quantum self-consistent approach Eqs.~(\ref{s3-eq-f}) and (\ref{s3-eq-current-monatomic}) (lines) and the quantum master equation (empty symbols)~\cite{Thingna12}. Since the quantum master equation is exact for any strength of anharmonicity we treat it as our benchmark to validate our approach. In Fig.~\ref{fig1} we see an excellent agreement between the two fundamentally different approaches for considerably high values of anharmonicity and for the entire temperature range. Inset of Fig.~\ref{fig1} shows the strong dependence of temperature on the transmission function, indicating the significant role anharmonicity plays in this system.
\subsection{Diatomic molecule}
\begin{figure}
\includegraphics[width=\columnwidth]{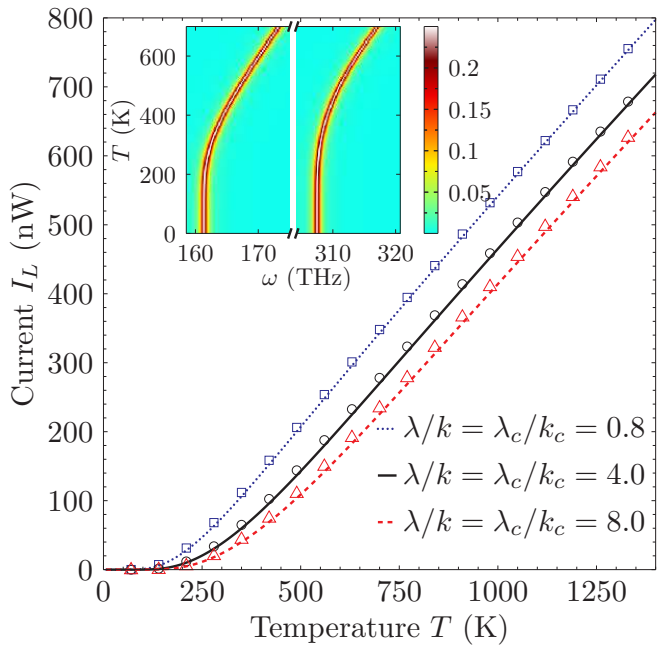}
\caption{\label{fig2}(Color online) Current $I_{L}$ as a function of temperature $T$ for various strengths of quartic interaction and quartic onsite potential in a diatomic molecule. Lines correspond to the quantum self-consistent approach and the empty symbols correspond to quantum master equation. Inset shows the temperature dependence of the transmission function for $\lambda/k=\lambda_c/k_c = 8$[$\textup{\AA}^2$u]$^{-1}$. All common parameters are: $m=$ 1u, $k=k_c$ 30.160meV/($\textup{\AA}^2$u), $\gamma=$ 0.92THz, $\omega_{c}=$ 0.92PHz, and $\Delta=0.05$. The left bath temperature $T_{L} = T\left(1+\Delta\right)$, whereas the right bath temperature $T_{R} = T\left(1-\Delta\right)$. $\lambda/k$ and $\lambda_c/k_c$ have dimensions of [$\textup{\AA}^2$u]$^{-1}$.}
\end{figure}

The Hamiltonian for the diatomic molecule is given by
\begin{eqnarray}
H_{S}=\frac{p^{2}_{1}}{2m}+\frac{p^{2}_{2}}{2m}&+&\frac{k}{2}(x^{2}_{1} +x^{2}_{2}) +\frac{k_{c}}{2}(x_{1} -x_{2})^{2} \nonumber \\
&+&\frac{\lambda}{4}(x^{4}_{1}+x^{4}_{2})+\frac{\lambda_{c}}{4}(x_{1} -x_{2})^{4},
\end{eqnarray}
where $\lambda$ and $\lambda_c$ are the strength of the anharmonic interaction potential and onsite potential respectively. The effective Hamiltonian according to QSCPT reads
\begin{equation}
H_{S}^{eff}=\frac{p^{2}_{1}}{2m}+\frac{p^{2}_{2}}{2m}+\frac{f}{2}(x^{2}_{1} +x^{2}_{2})+\frac{f_{c}}{2}(x_{1}-x_{2})^{2},
\end{equation}
where $f$ and $f_{c}$ are obtained by solving the following self-consistent nonlinear equations:
\begin{eqnarray}
\label{eq:fdi}
f&=&k+\frac{3\hbar\lambda}{4m}\left[\frac{1}{\Omega_{1}}\coth\left(\frac{\beta\hbar\Omega_{1}}{2}\right) \right. \nonumber \\
&& \left.+\frac{1}{\Omega_{2}} \coth\left(\frac{\beta\hbar\Omega_{2}}{2}\right)\right],\\
\label{eq:fcdi}
f_{c}&=&k_{c}+\frac{3\hbar\lambda_{c}}{m\Omega_2}\coth\left(\frac{\beta\hbar\Omega_{2}}{2}\right),
\end{eqnarray}
where $\Omega_{1}=\sqrt{f/m}$ and $\Omega_{2}=\sqrt{(2f_{c}+f)/m}$. Then the heat current across the diatomic molecule is given by
\begin{eqnarray}
I_{L}&=&\frac{1}{2\pi}\int_{0}^{\infty}d\omega\hbar\omega \mathrm{Tr}(G^{r}\Gamma_{L} G^{a}\Gamma_{R})(f_{L}-f_{R}) \nonumber \\
&=&\frac{2}{\pi}\int_{0}^{\infty}d\omega\hbar\omega\left|G^{r}_{12}(\omega)\right|^{2}J^2(\omega)(f_{L}-f_{R}),
\label{s3-eq-current-diatomic}
\end{eqnarray}
where
\begin{eqnarray}
G^{r}_{12}(\omega)&=&\frac{f_{c}}{\left[b^{2}(\omega)-J^{2}(\omega)-f_{c}^{2}+i2b(\omega)J(\omega)\right]},\nonumber \\
b(\omega)&=&m\omega^{2}-(f+f_{c}+\gamma m\omega_{c})-\frac{J(\omega)\omega}{\omega_{c}}.
\end{eqnarray}
Figure~\ref{fig2} shows a comparison between our quantum self-consistent approach Eqs.~(\ref{eq:fdi}), (\ref{eq:fcdi}), and (\ref{s3-eq-current-diatomic}) (lines) and the quantum master equation (empty symbols) for various combinations of the anharmonic parameters $\lambda$ and $\lambda_c$. The favorable agreement further validates our approach and the inset of Fig.~\ref{fig2} shows the temperature dependence of the transmission function indicating the strong role of anharmonicity.
\section{Thermal rectification in a two-segment model}\label{sec:4}
In order to show the strength of the quantum self-consistent approach we consider the stationary heat current across a chain consisting of two weakly coupled lattices,
\begin{equation} \label{s5-eq-Hamitonian}
H_{C} = H_{1}+\frac{k_{int}}{2} (x_{N/2+1}-x_{N/2})^2+H_{2}.
\end{equation}
The Hamiltonian for the left and right segments are given by
\begin{equation}
H_{1} = \sum^{N/2}_{n=1} \frac {p_n^2}{2m} +V_{1}(x_{n+1}-x_{n}) +U_{1}(x_{n})
\end{equation}
and
\begin{equation}
H_{2} = \sum^{N}_{n=N/2+1} \frac {p_n^2}{2m} +V_{2}(x_{n+1}-x_{n}) +U_{2}(x_{n}).
\end{equation}
Above $V_{1(2)}$ represents the interaction potential and $U_{1(2)}$ represents the onsite potential. The classical version of this model has been extensively studied to investigate the thermal rectification effect~\cite{Li_rev}. The occurrence of thermal rectification requires: i) Asymmetry and ii) Anharmonicity in the system. The model above exhibits spatial asymmetry due to the two segments having different parameters and anharmonicity due to the nonlinear interaction potentials. Hence, we choose the potentials of the two segments $a=1,2$ to take the form
\begin{equation} \label{s5-eq-int-potential}
V_{a}(x)=\frac{1}{2}k_{c,a}x^2+\frac{1}{4}\lambda_{c,a}x^4,
\end{equation}
and
\begin{equation} \label{s5-eq-on-potential}
U_{a}(x)=\frac{1}{2}k_{a}x^2+\frac{1}{4}\lambda_{a}x^4.
\end{equation}

\begin{figure}
\includegraphics[width=\columnwidth]{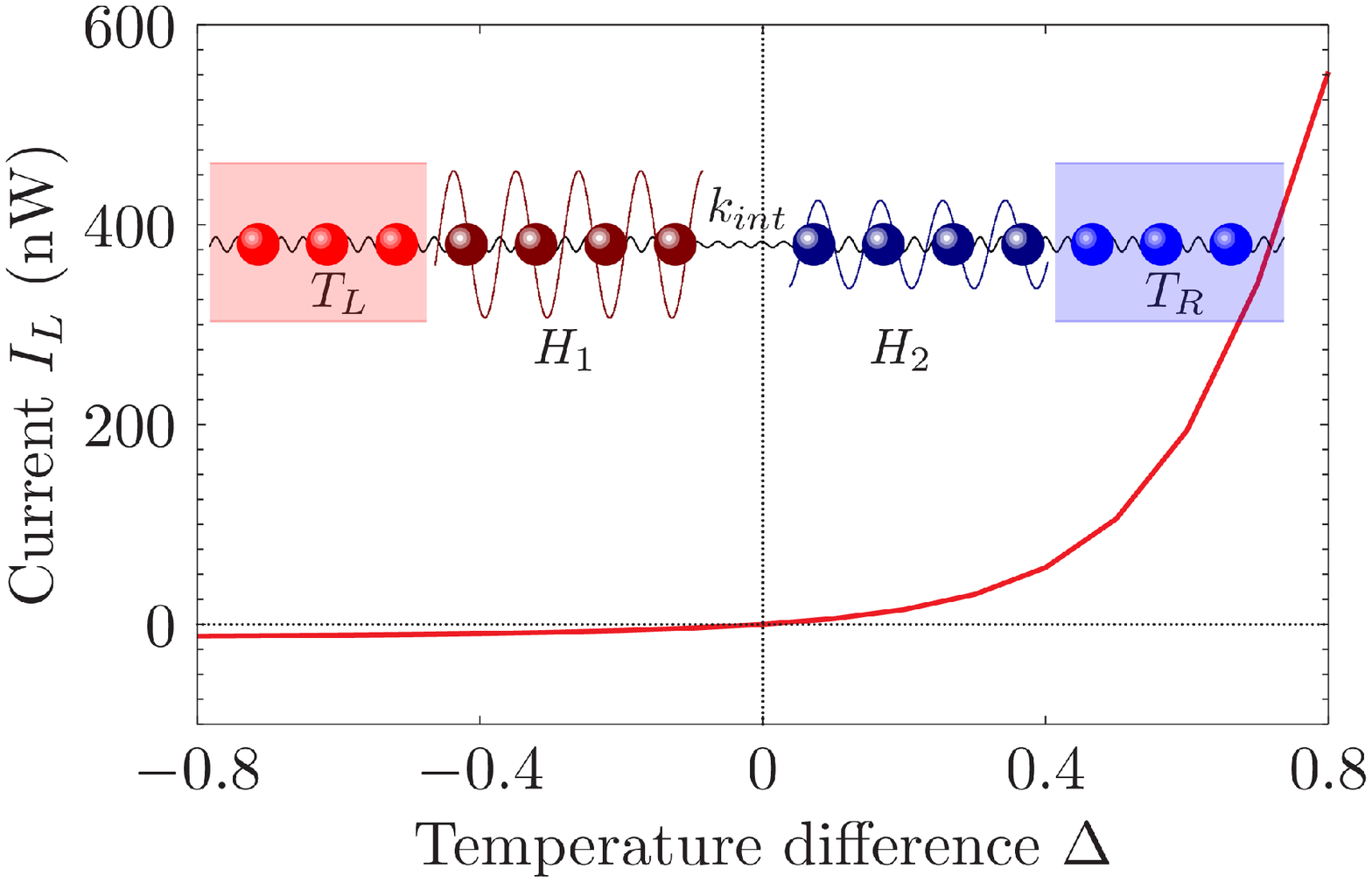}
\caption{\label{fig5}(Color online) Current $I_{L}$ as a function of relative temperature difference $\Delta$. Parameters used for the calculation are: $m=$ 1u, $k_{c,1}=k_{c,2}=k_1=k_2=$ 60.321meV/($\textup{\AA}^2$u), $2\lambda_{c,1}/k_{c,1}=\lambda_{c,2}/k_{c,2}=$ 2[$\textup{\AA}^2$u]$^{-1}$, $2\lambda_1/k_1=\lambda_2/k_2=$ 0.4[$\textup{\AA}^2$u]$^{-1}$, $k_{int}=$ 3.016meV/($\textup{\AA}^2$u), $\gamma=$ 0.92THz, $\omega_{c}=$ 0.92PHz, $T=$ 490K, and $N=8$. The left bath temperature $T_{L} = T\left(1+\Delta\right)$, whereas the right bath temperature $T_{R} = T\left(1-\Delta\right)$. Schematic shows the model considered to obtain the thermal rectification.}
\end{figure}

In order to apply the quantum self-consistent approach it is crucial that the system remains at approximately at one temperature $T$. Since the goal is to study thermal rectification, a far from linear response phenomenon, we choose the segment-segment coupling $k_{int}$ to be weak. This weak coupling between the two segments causes each segment to attain a temperature close to the bath it is connected to, i.e., the left segment $H_1$ attains the temperature $T_L$ and $H_2$ attains $T_R$. In accordance with our corroboration in Sec.~\ref{sec:3} we choose each segment to be weakly coupled to its respective bath. This allows us to safely apply our quantum self-consistent approach to each segment separately. According to QSCPT, the Hamiltonian $H_{1(2)}$ can thus be
approximated by the effective Hamiltonian $H_{1(2)}^{eff}$ that takes the form
\begin{equation}\label{s5-eq-H1eff}
H_{1}^{eff} = \sum^{N/2}_{n=1} \frac {p_n^2}{2m} +\frac {f_{c,1}}{2} (x_{n+1}-x_{n})^2 +\frac {f_{1}}{2}x_n^2,
\end{equation}
and
\begin{equation}\label{s5-eq-H2eff}
H_{2}^{eff} = \sum^{N}_{n=N/2+1} \frac {p_n^2}{2m} +\frac {f_{c,2}}{2} (x_{n+1}-x_{n})^2 +\frac {f_{2}}{2}x_n^2.
\end{equation}
The temperature of left and right heat bath is given by $T_{L(R)}=T(1\pm\Delta)$. Using the effective Hamiltonian Eqs.~(\ref{s5-eq-H1eff}) and (\ref{s5-eq-H2eff}) in Eq.~(\ref{s5-eq-Hamitonian}) we can obtain the heat current through the system using the Landauer-like formula Eq.~(\ref{s2-eq-current}). Figure~\ref{fig5} shows the heat current as a function of the dimensionless temperature difference $\Delta$. The sign of $\Delta$ indicates the direction of the current, i.e., $\Delta>0$ corresponds to a current from the left lead to the right lead and vice versa. We find that the heat current is substantially larger for $\Delta>0$ than that for $\Delta<0$ leading to a large thermal rectification ratio $R= |I_L(\Delta)-I_L(-\Delta)|/{\rm max}\{I_L(\Delta),I_L(-\Delta)\}$, where $I_L(\Delta)$ is the current evaluated at a fixed value of temperature difference $\Delta$. The maximum rectification we achieve is $\approx 98\%$ and to the best of our knowledge this value far exceeds the values obtained in the quantum regime \cite{WuPRL09, WuPRE09}.
\section{Conclusion}\label{sec:5}
We develop a quantum self-consistent approach to study thermal transport across model-independent anharmonic systems. The key idea is to renormalize the anharmonic system to an effective harmonic one through a nonperturbative self-consistent approach. The effective Hamiltonian helps us utilize the nonequilibrium Green's function machinery, that is exact for Harmonic systems, in order to evaluate the heat current. In case of strong anharmonic systems we corroborate our approach with the master equation based formulation and find excellent agreement for the entire temperature range for mono- and di-atomic systems. Moreover, we also tackle an interesting two segment anharmonic model consisting of $8$ particles that is well beyond the reach of master equation based formulations. The two segment model exhibits a significantly large rectification ratio in the quantum regime, which is due to the strong temperature dependent phonon bands that overlap unequally leading to the large rectification ratio \cite{Hu_SCPT06}.

Overall, the quantum self-consistent approach is highly efficient and can be extended to higher dimensions incorporating effects of mass disorder \cite{Chaudhuri10}. Also complicated anharmonic potentials like the onsite Morse potential \cite{Hu_SCPT06,He09} could be handled within this approach making it highly versatile. However, one should bear in mind that the approach in its present form has two limitations. Firstly, since we apply the canonical average to the system at approximately an average temperature, the approach cannot be applied to study homogeneous systems under a large temperature gradient. Although inhomogeneous systems such as the two segment model illustrated in Sec.~\ref{sec:4} can be easily studied. Secondly, even though anharmonicity can be exactly captured within this approach it fails to capture the diffusive behavior of systems. In other words the phonon mean-free path within this formulation is infinite. Thus, the approach captures essential physics of only those systems that are shorter than its actual phonon mean-free path making it relevant to the field of nano-device engineering.

\begin{acknowledgments}
We acknowledge the helpful discussions with Sahin Buyukdagli, Lifa Zhang and late Prof. Bambi Hu. D. H. is surported by NSFC of China (Grant Nos. 11105112 and 11335006) and NSF of Fujian Province (No. 2016J01036). J.-S. W. acknowledges support from an FRC grant R-144-000-343-112.
\end{acknowledgments}
%
\end{document}